# Carbon-Taxed Transformers: A Green Compression Pipeline for Overgrown Language Models

AJMAIN INQIAD ALAM, University of Saskatchewan, Canada
PALASH ROY, University of Saskatchewan, Canada
CHANCHAL K. ROY, University of Saskatchewan, Canada
BANANI ROY, University of Saskatchewan, Canada
KEVIN A. SCHNEIDER, University of Saskatchewan, Canada

The accelerating adoption of Large Language Models (LLMs) in software engineering (SE) has brought with it a silent crisis: unsustainable computational cost. While these models demonstrate remarkable capabilities in different SE tasks, they are unmanageably large, slow to deploy, memory-intensive, and carbon-heavy. This reality threatens not only the scalability and accessibility of AI-powered SE, but also its long-term environmental sustainability. The research challenge is clear: we must go beyond accuracy and address efficiency and environmental cost as first-class design constraints. To meet this challenge, we introduce Carbon-Taxed Transformers (CTT), a systematic multi-architectural compression principled pipeline ordering inspired by economic carbon taxation principles. Drawing from the economic concept of carbon pricing, CTT operationalizes a computational carbon tax that penalizes architectural inefficiencies and rewards deployment-ready compression. We evaluate CTT across three core SE tasks: code clone detection, code summarization, and code generation, with models spanning encoder-only, encoder-decoder, and decoder-only architecture. Our results show that CTT delivers on inference: (1) up to 49× memory reduction, (2) time reduction up to 8-10× for clone detection, up to 3× for summarization, and 4–7× for generation, (3) up to 81% reduction in $CO_2$ emissions and (4) CTT retains around 98% accuracy on clone detection, around 89% on summarization, and up to 91% (textual metrics) and 68% (pass@1) for generation. Two ablation studies show that pipeline ordering and individual component contributions are both essential, providing empirical justification for CTT's design and effectiveness. This work establishes a viable path toward responsible AI in SE through aggressive yet performance-preserving compression.

CCS Concepts: • **Computing methodologies** → **Neural networks**; *Natural language processing*; • **Software and its engineering** → **Software development techniques**; *Software notations and tools*; • **Hardware** → **Power and energy**.

Additional Key Words and Phrases: Model Compression, Green Computing, LLMs, Energy-Efficient Inference

## 1 Introduction

Large Language Models (LLMs) have rapidly advanced the state-of-the-art (SOTA) in software engineering (SE) tasks [1, 6, 59, 64, 65]. In particular, code-oriented LLMs such as OpenAI Codex

Authors' Contact Information: Ajmain Inqiad Alam, University of Saskatchewan, Saskatoon, Canada, ajmain.alam@usask.ca; Palash Roy, University of Saskatchewan, Saskatoon, Canada, palash.roy@usask.ca; Chanchal K. Roy, University of Saskatchewan, Saskatoon, Canada, chanchal.roy@usask.ca; Banani Roy, University of Saskatchewan, Saskatoon, Canada, banani.roy@usask.ca; Kevin A. Schneider, University of Saskatchewan, Saskatoon, Canada, kevin.schneider@usask.ca.

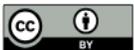

and GitHub Copilot [8] and DeepMind's AlphaCode [29] demonstrate that generative AI can assist with code synthesis, debugging, and other development activities at a high level. These foundation models, often containing tens or hundreds of billions of parameters, are capable of producing functionally correct code and even solving competitive programming problems [29]. Early studies report promising productivity gains when such AI pair programmers are introduced into real developer workflows, for example, a controlled experiment found that developers using GitHub Copilot completed a coding task 55% faster than those without it [39]. This potential of LLMs in SE has fueled a new paradigm of AI-assisted programming, wherein neural models become integral collaborators in software development.

Although LLMs demonstrate remarkable performance, they are very large with billions (B) of parameters, incur high operational costs, consume substantial computational memory, require specialized hardware, and suffer from low inference speed. For instance, GPT-3 (175B parameters) requires 325 GB of GPU memory to load model weights, necessitating at least five A100 GPUs (each with 80 GB of memory) along with advanced parallelism techniques to function effectively during inference [48]. While LLMs such as GPT offer APIs for accessibility, utilizing these APIs can significantly elevate operational costs. Even open-source models like CodeGen, which have 16B parameters, require 72 GB of memory. The large memory and energy demands of these models also lead to a substantial carbon footprint [52]. Also, training large NLP models can emit significant quantities of $CO_2$ [38]. Training a model like GPT-3 emits three times more $CO_2$ than a jet plane from San Francisco to New York [38]. Moreover, inference has become a dominant contributor to long-term energy consumption in deployed AI systems. AWS and NVIDIA estimate that inference accounts for 80–90% of total ML computing, and approx. 60% of total ML energy use at the inference in Google [33, 37, 38]. The trend toward ever-larger models has negative concerns around sustainability and accessibility in SE.

One promising direction to address these challenges is model compression, an active area of research in both NLP and SE. The goal is to drastically reduce model size and computation without significant loss of accuracy. Techniques such as knowledge distillation (KD) [24, 44], quantization [47, 69], and network pruning [35] have each shown success on general language models. For instance, DistilBERT distilled a BERT model to nearly half its depth and achieved > 95% of the original accuracy [44], and TinyBERT went further to < 30% of BERT's parameters with minimal performance drop [24]. Weight quantization can shrink models by using low-bit (8-bit or even 4-bit) representations; Q8BERT demonstrated 8-bit BERT with negligible accuracy loss [69], and subsequent work has pushed to ultra-low 4-bit precision for Transformers [47]. Pruning studies have found that large portions of Transformer weights can be removed while preserving most accuracy [35].

In the SE domain, initial efforts mirror these trends. Su et al. [54] distilled a GPT model for code summarization and found the small student model can match a GPT-3.5 teacher on that task, and Wei et al. [63] achieved 30% memory reduction on a 16B parameter code model via 8-bit quantization with negligible impact on code generation quality. Neural architecture search (NAS) offers another axis of optimization, automatically discovering efficient model architectures. Recent NAS approaches for Transformers have yielded designs (the Primer architecture [51]) that outperform human-designed ones in speed/accuracy trade-offs. Despite this progress, important gaps remain. Prior works [23, 44, 49, 50] typically apply a single compression technique (distill or quantize or prune), often targeting a specific task. Little attention has been paid to combining techniques for maximum efficiency gains, and to our knowledge no existing approach holistically tailors compression specifically for the diverse workloads and architectures in SE.

To address the rising computational and environmental costs of LLMs, we propose the **Carbon-Taxed Transformers (CTT)** pipeline, a systematic multi-architectural compression pipeline that operationalizes economic carbon taxation principles for model optimization. This tax is not a literal fee but a set of algorithmic constraints applied during model compression, forcing a trade-off between performance and efficiency. CTT addresses this with three key innovations: (1) *architectural generalization* across encoder-only, decoder-only, and encoder-decoder models, (2) *principled pipeline ordering* that systematically sequences NAS→pruning→quantization→KD to maximize deployment compatibility, and (3) *budget-aware optimization* that enforces hard constraints on inference latency, memory, and $CO_2$ emissions. CTT is evaluated across three key SE tasks, each tied to a distinct Transformer type: encoder-only (code clone detection (CCD)), decoder-only (code generation), and encoder–decoder (code summarization), ensuring broad architectural coverage and validating generality. Unlike prior pipelines [43, 49, 50], which are limited to single architectures and do not examine how ordering impacts results under deployment budgets, CTT provides a unified framework that demonstrates consistent improvements across the evaluated tasks and architectures while maintaining competitive performance. While Parameter-Efficient Fine-Tuning (PEFT) methods reduce fine-tuning costs by updating a small number of parameters, they retain the full pre-trained model during inference [4, 12, 20], and thus fall outside the scope of our work. Our design yields models that are compact, fast, and environmentally aligned. By reducing complexity, energy consumption, and emissions without sacrificing task effectiveness, CTT imposes a computational *carbon tax* on overparameterized transformers, producing models that retain competitive accuracy at a fraction of the environmental and computational cost. This focus is especially important given that inference accounts for the majority of energy use and $CO_2$ emissions, often exceeding 90% in real-world deployments [38]. To guide our investigation, we pose the following research questions:

**RQ1:** *How much can a compressed student model preserve the performance of its teacher across diverse SE tasks?*
**RQ2:** *How does the inference speed of a small student model compare to its teacher model on CPU and GPU?*
**RQ3:** *To what extent can model compression reduce the memory demands of LLMs during inference on CPU and GPU?*

In this work, we: (1) propose CTT, a systematic multi-architectural compression pipeline integrating NAS, structured pruning, quantization, and KD with empirically validated principled ordering, (2) demonstrate 4–10× inference speedups and up to 49× memory reduction across encoder-only, decoder-only, and encoder–decoder models for three core SE tasks, (3) show that CTT maintains strong performance despite aggressive compression (98% accuracy retention in clone detection, 89% in summarization, and 91% in generation), and (4) provide an environmental impact analysis showing consistent $CO_2$ emission reductions of up to 81% during inference across all tasks.

The remainder of the paper is organized as follows: Section 2 outlines our methodology, followed by the experimental setup in Section 3. Results are presented in Section 4, with an ablation study and a discussion in Section 5 and 6 respectively. Section 7 covers threats and mitigations, related work is reviewed in Section 8 and conclusions are drawn in Section 9. Section 10 presents the acknowledgments, and Section 11 has data availability.

## 2 Methodology

This section outlines the core methodology behind our compression pipeline. We begin with a high-level overview of the compression process, followed by detailed descriptions of each stage in the pipeline. Our compression approach is structured into *three sequential stages*: (1) structural reduction, where we impose a tax on architectural complexity by eliminating redundant components; (2)

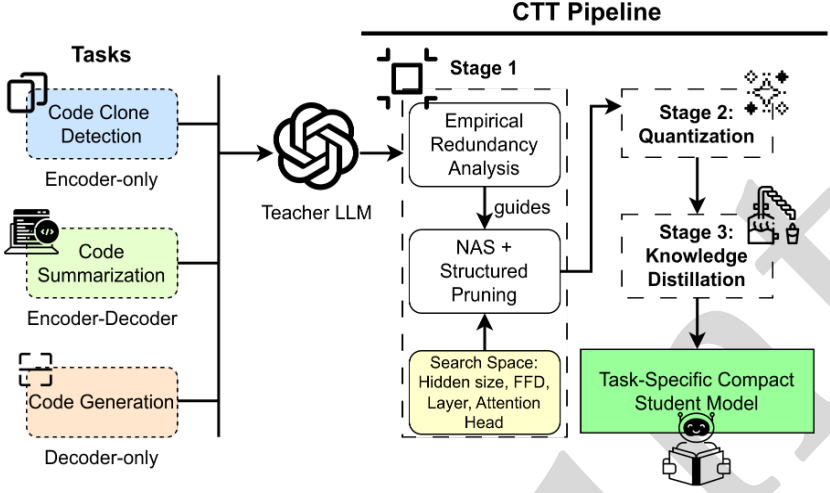

Fig. 1. Carbon-Taxed Transformers pipeline

---

**Algorithm 1:** CTT Pipeline for Student Model Training

---

**Require:** Teacher model $T$, NAS configuration space $C$, training dataset $D$
**Ensure:** Compressed and optimized student model $S_{\text{final}}$
1: **Stage 1: Structural Reduction**
2: 　$S_{\text{nas}} \leftarrow$ NAS_search($C, D$) // Search for an initial compact architecture
3: 　**for** each component ∈ {layers, attention heads, hidden size, feedforward size} **do**
4: 　　Iteratively reduce component size in $S_{\text{nas}}$
5: 　　Evaluate performance after each reduction on $D$
6: 　　**if** performance drop is negligible **then**
7: 　　　Accept the reduction
8: 　　**else**
9: 　　　Revert to previous configuration
10: 　**end for**
11: 　$S_{\text{pruned}} \leftarrow$ refined architecture after all accepted reductions
12: **Stage 2: Quantization-aware Initialization**
13: 　$S_{\text{quant}} \leftarrow$ quantize($S_{\text{pruned}}$) // Convert weights to low-precision format before training
14: **Stage 3: Knowledge Distillation**
15: 　Train $S_{\text{quant}}$ using KD loss with teacher model $T$ // Align student outputs with teacher predictions
16: **return** $S_{\text{final}} \leftarrow S_{\text{quant}}$

---

quantization-aware initialization, which imposes a tax on numerical precision to lower memory and compute requirements; and (3) a performance rebate through KD, which transfers performance-relevant knowledge from the teacher to the student model. The full pipeline is illustrated in Figure 1, and Algorithm 1 outlines the training workflow.

In designing the CTT pipeline, we focus on compressing four key architectural parameters rather than tuning training hyperparameters. We target the number of hidden layers, attention heads, hidden dimensions, and feedforward dimension (FFD) size. These four components were selected because they map directly to common compression targets and each of them significantly impacts memory, latency, and model size. Unlike abstract metrics or task-agnostic statistics, these measures provide actionable insights for optimizing architecture with minimal loss in performance.

## 2.1 Stage 1: Structural Reduction

Transformer-based language models have greatly improved SE tasks [8, 29], and are commonly built using one of three architectures: decoder-only [27, 36], encoder-only [13, 32], or encoder-decoder [1, 61]. Although these architectures are powerful, they are often designed with the same number of layers throughout, even though not all layers contribute equally to learning. This leads to models that are larger and slower than necessary. To address this, we begin by analyzing how different parts of the model behave during inference. We select three SE tasks CCD, summarization, and generation, each using a different type of Transformer. This helps ensure that our compression approach is based on real model behavior rather than assumptions.

*2.1.1 Empirical Analysis of Layer Redundancy.* As our focused is on four architectural components, we have used following metrics: (1) hidden state norm, the $\ell_2$ norm of token-wise hidden representations per layer, scaled by hidden dimension, capturing the strength of intermediate representations; (2) feedforward norm, computed after each feedforward block, which reflects the layer's transformation intensity; (3) attention head score, the average attention activation across all heads in a layer, measuring contextual focus and redundancy; and (4) depth prior, $1/(l+1)$, where $l$ is the zero-indexed layer number, encoding the intuition that deeper layers often have diminishing returns.

*Code Clone Detection.* In Figure 2 for CCD, we observe that hidden state norms and feedforward norms rise sharply through early layers but plateau beyond the midpoint (layer 8–10). Attention head scores, which ideally reflect diverse context aggregation, become uniform in later layers (avg. 0.077), indicating redundancy. In terms of efficiency, while FLOPs increase linearly by layer, representation strength does not, particularly beyond layers 8–10, showing high compute with low return. This suggests that deeper layers act as near-identity mappings with high FLOPs but minimal representational gain, a signal for safe reduction.

*Code Summarization.* In Figure 2 for code summarization models, the encoder's hidden state norm follows a similar trajectory: increasing up to a peak and declining or saturating after mid-depth. This highlights that early encoder layers carry the semantic compression burden, while the final layers mostly serve as relays. Attention scores again show little variation in deeper layers, implying converged and redundant attention patterns. Interestingly, decoder layers also show flattening in both norm and attention metrics, suggesting that only a fraction of decoder depth is critical for meaningful output.

*Code Generation.* In Figure 2 for code generation, trends were similarly pronounced. Hidden state norms peak around layers 10–12 and then plateau. Attention heads in deeper layers show less focus diversity, converging toward uniformity. This reflects their autoregressive nature: short-term dependencies are captured early, while deeper layers offer diminishing utility (a classic inefficiency). These findings provide a natural cut-off point.

Our empirical analysis reveals a consistent pattern: early and mid-depth layers contribute most to semantic representation, while deeper layers are often computationally expensive yet offer limited functional value. Metrics such as hidden state norms, feedforward activity, and attention head diversity plateau or decline in later layers, indicating diminishing returns on compute and exposing substantial architectural redundancy. While more sophisticated alternatives exist, such as gradient-based importance scores [45], applying them across our diverse range of target models and tasks would be considerably more resource-intensive [58] and prone to instability [14]. By contrast, our lightweight diagnostics provide interpretable signals that scale across Transformer families and SE tasks. Based on these findings, we now have a clear understanding of where structural reduction can be applied most effectively. Although prior work has employed NAS to discover efficient student models [11, 50, 56], such methods often rely on proxy objectives like accuracy or latency and may overlook internal inefficiencies. Consequently, NAS can yield architectures that are over-parameterized, structurally irregular, or ill-suited for quantization and real-world deployment [5, 30]. To address these shortcomings, we adopt a hybrid strategy that combines NAS with structured pruning. NAS is used to generate an initial compact candidate, after which we iteratively prune redundant components from selected four architectural component based on performance. This approach balances architectural exploration with interpretability and fine-grained control, enabling us to produce student models that maintain structural fidelity to the teacher (beneficial for KD [60]), while improving efficiency, quantization compatibility, and generalizability across diverse SE tasks.

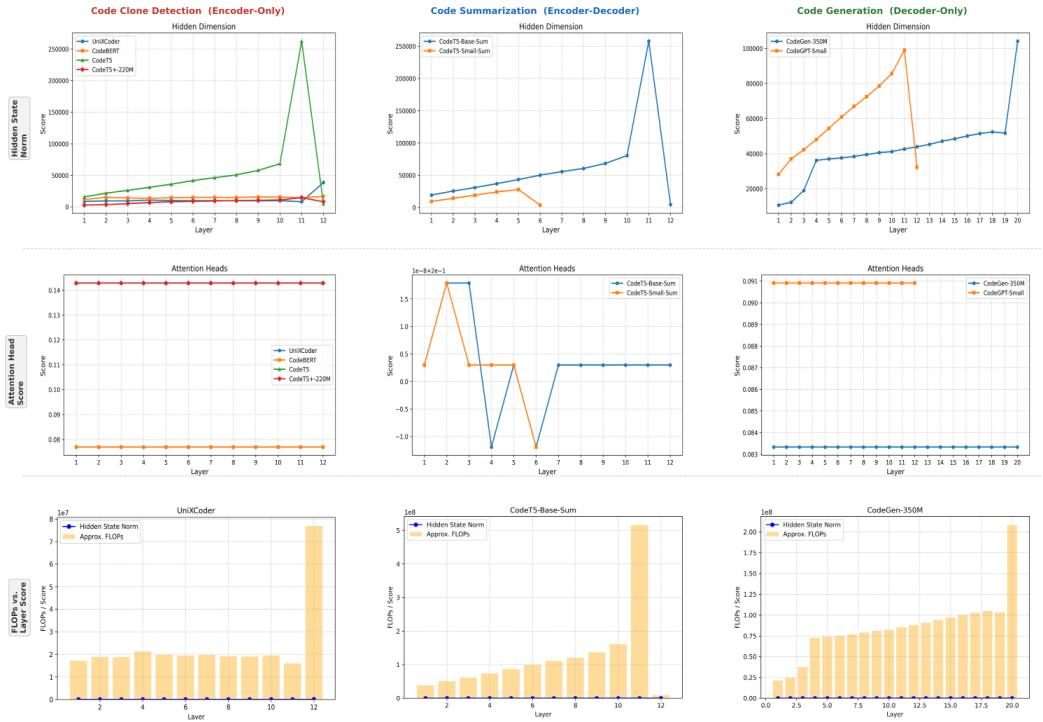

Fig. 2. Empirical analysis of layer redundancy across the three SE tasks. Row 1 (Hidden State Norm): representation strength rises through early layers then plateaus or saturates beyond mid-depth, indicating diminishing returns in deeper layers. Row 2 (Attention Head Score): scores converge toward uniformity across layers, signalling redundant contextual aggregation; feedforward norms follow the same pattern and are omitted for brevity. Row 3 (FLOPs vs. Layer Score): approximate FLOPs per layer (bars) against hidden-state norm score (dots) for the primary teacher per task; later layers incur disproportionately higher compute cost relative to their marginal representational contribution. These three signals collectively identify safe reduction cut-off points that motivate the NAS and structured pruning stages in CTT.

In the final phase of this stage, we begin by selecting a teacher language model and applying NAS over a predefined configuration space $C$. NAS identifies an initial compact student model by optimizing validation performance on dataset $D$ while reducing computational cost. We minimize validation loss subject to latency $\leq T_{\max}$ and peak memory $\leq M_{\max}$ on the target GPU/CPU.

Infeasible candidates are discarded; among feasible ones we select the best validation loss. However, the resulting architecture often still contains redundant components. To address this, we refine the NAS-derived model through structured pruning. At each step, we iteratively reduce capacity and evaluate performance, retaining only those changes that lead to negligible or minor accuracy loss. This process operationalizes our carbon tax: we determine whether the efficiency gain from a reduction justifies the performance tax it imposes. We do not impose too much reduction on the model, which would degrade performance, nor too little, which would result in retaining excessive architectural overhead. In doing so, this stage enforces a balanced reduction, tightening the model's footprint without compromising its utility. For instance, in the CCD task using UniXCoder as the teacher, NAS produced a student with 10 layers, 8 attention heads, FFD size 96, and hidden size 32, which was further pruned to 8 layers, 4 heads, FFD size 8, and hidden size 8 without significant performance degradation.

### 2.2 Stage 2: Quantization-aware initialization

The pruned student model is subsequently subjected to quantization prior to training to further compress its memory footprint and improve computational efficiency. This stage imposes a 'tax on precision' by using lower-precision numbers, while preserving performance For example, A 32-bit GPT model taking 10 GB of memory can be quantized to 8-bit and shrink to 2.5 GB, enabling faster inference with minimal accuracy loss. By quantizing before training, we force the model to learn under this numerical tax from the outset. Performing quantization before training allows the model to be exposed to quantization-induced numerical noise during optimization, enabling it to better adapt and maintain performance during inference under low-precision constraints. This design choice mitigates the accuracy degradation often observed when quantization is applied only after full-precision training [23]. The resulting quantized model is denoted as $S_{\text{quant}}$ and serves as the initialization for the final KD step.

### 2.3 Stage 3: Knowledge Distillation

After the student model has been taxed for its architectural and numerical complexity, this final stage acts as a performance rebate. The quantized student model is trained using KD, a technique that allows a smaller model to learn from a larger, more accurate teacher model. Instead of relying only on hard labels, the student is trained to match the teacher's output distribution, which contains more detailed information about how the teacher interprets each input. For example, rather than simply predicting the correct token, the student learns to imitate the teacher's probability distribution over all possible tokens, capturing how confident the teacher is about different predictions. We implement this by using a distillation loss that minimizes the difference between the soft logits of the teacher model $T$ and those of the student model $S_{\text{quant}}$ on the same input samples from dataset $D$. This process helps the student reclaim performance that was paid during the taxation stages of structural reduction and quantization. The final result is an optimized student model, denoted as $S_{\text{final}}$, that remains compact and efficient while maintaining high task performance.

To summarize, each stage of the CTT pipeline plays a distinct but complementary role in achieving efficient and environmentally conscious model compression. The "tax-and-rebate" terminology is a conceptual framing that directly corresponds to the sequence of operations described in previous sections. The *tax* is imposed through three capacity-reducing transformations—Neural Architecture Search (NAS), structured pruning, and quantization—each constraining the student model's representational capacity in a different way. NAS provides the foundational architecture by identifying compact, task-aligned configurations under latency and memory constraints, removing large-scale overparameterization early in the process by reducing layers, attention heads, and width dimensions with low contribution. However, NAS alone may miss fine-grained inefficiencies; structured pruning therefore refines the architecture by eliminating residual redundant components in an interpretable and performance-aware manner. Quantization applies an additional tax by lowering numerical precision, significantly reducing memory and computational requirements with minimal impact on accuracy. Together, these stages constitute the computational "carbon tax," shrinking the model's architectural and numerical footprint in a controlled manner. The *rebate* is implemented through KD, which restores task performance by transferring semantic knowledge from the teacher to the compressed student without reintroducing the removed structural or numerical complexity, thereby improving how efficiently the reduced capacity is utilized. Overall, this principled NAS→pruning→quantization→KD pipeline produces compact, efficient, and deployment-ready models, aligning performance objectives with sustainability goals.

## 3 Experimental Setup

This section details the experimental setup used to implement. To ensure that our design decisions are broadly applicable, we apply this methodology across three widely used Transformer architectures: encoder-only, decoder-only, and encoder-decoder. We selected clone detection, code summarization, and code generation as our three SE tasks because they are established, widely used, and collectively span all three major Transformer architectures, covering distinct inference behaviors that are directly relevant to evaluating CTT. Specifically, clone detection is a fast, discriminative task with minimal sequential dependency; summarization is a sequence-to-sequence problem with higher attention cost due to cross-encoder interactions; and code generation is autoregressive and the most compute-intensive of the three, making each task a meaningful stress test for a different efficiency bottleneck. Encoder-only models excel at discriminative tasks (clone detection), Encoder-decoder models for sequence-to-sequence tasks (code summarization) and Decoder-only models for generative tasks (code generation). Regarding the model selection, our primary goal was to demonstrate generalizability and effectiveness of CTT pipeline on established models which are used and validated in practical scenarios. These models were chosen for architectural representativeness, established performance, and community adoption in respective domains which is common in SE-literature [49, 50]. We also included foundational models that are frequently used as established baselines, and latest compression techniques like Avatar, which allow us to situate our performance gains within existing research. Furthermore, hardware constraints guided our choices to go beyond 1.5B, as we had to manage GPU load in shared environment where larger models lead to device out of memory issues. All student models were fine-tuned using AdamW, with a softened KL loss for KD. All code details and data are shared in the replication package.

### 3.1 Hardware Environment

All training and GPU-related-evaluation were conducted on a server equipped with Ubuntu 22.04.5 LTS, an Intel(R) Xeon(R) Platinum 8356H CPU @ 3.90GHz, 3.0 TiB RAM, and single NVIDIA A100 80GB PCIe GPU. To run CPU-related evaluation, we used a desktop machine running Ubuntu 16.04 LTS, with an Intel(R) Core i7-2600 CPU @ 3.4GHz and 12 GiB RAM. We include an i7-2600 to reflect legacy CI/build agents and developer laptops still common in industry; CTT specifically targets deployability on constrained hosts. GPU results are included for completeness.

### 3.2 Neural Architecture Search Methodology

We have employed a NAS methodology inspired by prior literature [49, 57]. We defined a discrete search space comprising: Number of transformer layers ∈ {2 to 12}, Number of attention heads ∈ {2, 4, 6, 8, 10, 12}, Hidden size is computed as the product of attention head and head dimension ∈ {2, 4, 8, 16, 32, 64, 96, 128} and FFD size ∈ {4 to 3072}. The algorithm employed for NAS utilized the following hyperparameters: a population size of 50, a total of 50 generations, a crossover rate of 0.7, and a mutation rate of 0.1. At each generation, the top two candidate architectures were preserved (elitism), while subsequent candidates were generated through crossover and mutation operations applied to the top four architectures. This iterative process continued for 50 generations, after which the most performant architecture was selected for subsequent structural pruning. NAS enforces latency and memory budgets as hard feasibility constraints during the search process: candidate architectures exceeding either limit are discarded prior to ranking, while the remaining feasible candidates are evaluated solely based on validation loss. Tightening these constraints narrows the feasible search space without altering the selection objective, leading NAS to shift toward smaller configurations in a gradual and predictable manner, as long as architectures with sufficient representational capacity remain. However, when constraints become overly restrictive and eliminate all viable candidates, performance degradation may occur due to insufficient model capacity. To mitigate this, the subsequent structured pruning stage provides an additional safeguard by reverting reductions that introduce unacceptable validation loss, ensuring stability across the pipeline. The task-dependent architecture sizes reported in Table 1 reflect this constraint-driven behavior: clone detection supports significantly smaller architectures due to its discriminative nature, whereas summarization and generation stabilize at relatively larger configurations owing to their sequence-to-sequence and autoregressive complexity.

### 3.3 Code Clone Detection Task

***Dataset.*** For the CCD task, we employ the BigCloneBench dataset [55]. We select BigCloneBench for two key reasons: (1) its large scale and richness, and (2) its status as a standard benchmark ensures comparability. BigCloneBench contains labeled Java method pairs, identifying whether each pair represents a code clone (i.e., functionally similar despite possible syntactic differences) or not. In our setup, the dataset has a training set of over 45,000 examples and separate validation and test sets of 4,000 examples each.

***Training and Evaluation Strategy.*** The architectural details for this task are detailed in Table 1. The student model is trained using logits generated by fine-tuned encoder based teacher model: UniXCoder [17]. The model learns only by mimicking the teacher's output. We evaluate model performance on the CCD task using precision and recall. Additionally, we report inference latency, memory usage and inference $CO_2$ emission analysis. In this evaluation, CodeBERT [13], CodeT5 [62], CodeT5+-220M [61], Compressor [50] and Avatar [49] are included as comparative baselines. For Avatar, we have selected one of the high performing architectures from the search. We have selected encoder state of UniXCoder as the teacher because it consists encoder-only architecture, which aligns well with the target architecture space explored during compression.

### 3.4 Code Summarization Task

***Dataset.*** For the code summarization task, we use the CodeSearchNet summarization dataset [22]. We select CodeSearchNet as: (1) it offers a substantial number of annotated examples, and (2) it is a widely adopted benchmark [18, 68], supporting reproducibility and comparisons with prior work [62]. CodeSearchNet contains source code snippets paired with natural language documentation across several programming languages. In this work, we use the Java subset. Our experimental setup includes 164,923 training examples, 5,183 for validation, and 10,955 for testing.

***Training and Evaluation Strategy.*** The architectural details are presented in Table 1. For this task, we use CodeT5-base-multi-sum [62] as the teacher model. We selected CodeT5-base-multi-sum as the teacher as it balances between representational strength and practical scale and is widely recognized as a strong encoder-decoder baseline. During experimentation, we attempted to include CodeT5+ (770M–16B) but encountered repeated AssertionError crashes (also reported on HuggingFace and GitHub [40, 41]). As these issues remain unresolved, we excluded CodeT5+ while noting that CTT remains compatible with stronger encoder–decoder teachers once stable. However, unlike the CCD setting, the teacher's performance on this task was relatively weak, achieving only 19.65 BLEU score during fine-tuning. Training the student model solely based on such soft predictions from the teacher resulted in poor generalization and significantly lower BLEU scores in early experiments. To address this, we adopt a modified strategy: the student model is trained using both the ground-truth supervision and the logits of the teacher model used in KD loss function. This approach stabilizes training and ensures that the student benefits from both high-quality supervision and the representational guidance of the teacher. Model performance is evaluated using the BLEU, METEOR, and ROUGE-L F1, alongside inference latency and memory usage on both CPU and GPU. In addition, we include an analysis of $CO_2$ emissions during inference, providing an environmental perspective on model efficiency. In this evaluation, CodeT5-base-multi-sum, CodeT5-small [62], and our CTT-based model are compared.

### 3.5 Code Generation Task

***Dataset.*** For the code generation task, we use the CodeSearchNet generation challenge [22] which contains function-level code snippets paired with their corresponding docstrings dataset, specifically the Python subset, which is widely used for evaluating code generation models. For our experiments, we use 70,000 examples for training, 2,500 examples for validation, and 5,000 examples for testing. Since CodeSearchNet does not include executable test cases, it cannot directly support functional correctness evaluation (pass@1). To address this, we also incorporate the Google Research MBPP (Mostly Basic Python Problems) dataset [2], using its sanitized subset, which provides problem descriptions alongside reference implementations and corresponding unit tests. This enables direct measurement of functional correctness. The sanitized MBPP subset includes 120 training samples, 43 validation samples, and 257 test samples.

***Training and Evaluation Strategy.*** The architectural configuration is detailed in Table 1. In this task, we use CodeGen-350M [36] and Qwen Coder 2.5 1.5B Instruct [21, 67] as teacher models, selected for their decoder-only architecture, balance latest and established models, supported within our GPU and CPU resources. Initial experiments showed that relying solely on the teacher's outputs led to degraded student performance. To overcome this, we adopted a hybrid training strategy combining ground-truth supervision with teacher logits in the KD loss with higher emphasis on KD loss and temperature of 2. This technique leverages both accurate supervision and representational guidance, helping the student model maintain fluency while reducing dependency on a weak teacher. We assess model performance using BLEU, CodeBLEU, and pass@1. BLEU provides a surface-level measure of similarity to reference solutions, while CodeBLEU extends this with code-aware features such as syntax and data flow. Pass@1 evaluates functional correctness by testing whether the top generated program executes successfully, reflecting practical utility beyond text similarity. To evaluate the practicality of deployment, we also measure inference time and memory consumption on both CPU and GPU platforms. Finally, we report $CO_2$ emissions during inference, extending our analysis to include environmental sustainability. The evaluation compares four models: Codegen-350M, Qwen Coder 2.5 1.5B Instruct, and our compressed CTT-based models which compressed both the CodeGen and Qwen Coder.

## 4 Case Studies and Experimental Insights

### 4.1 RQ1: Performance comparison

Balancing compactness and capability in neural models is like walking a tightrope. Yet, not all reduction needs to be catastrophic. In this RQ, we dissect that balance, evaluating how much performance can be preserved when aggressively shrinking a large teacher model into a compact student.

Table 1. Comparison of Teacher and CTT-based Student Architectures Across Tasks

| Task | Model Family | Role | Model Size | Trainable Params | Attention Heads | FFD Size | Hidden Dim. | Layers |
|---|---|---|---|---|---|---|---|---|
| Clone Detection | CodeBERT | Teacher | 481 MB | 125M | 12 | 3072 | 768 | 12 |
| | UniXCoder | Teacher | 504 MB | 125M | 12 | 3072 | 768 | 12 |
| | CodeT5 | | 892 MB | 220M | 12 | 3072 | 768 | 12 |
| | CodeT5+ | | 446 MB | 220M | 12 | 3072 | 768 | 12 |
| | Compressor | Baseline | 3.1 MB | ~0.75M | 8 | 64 | 96 | 12 |
| | Avatar | Baseline | 5.5 MB | ~1.4M | 1 | 1508 | 36 | 3 |
| | CTT-based Model | Student | **1.6 MB** | **~0.42M** | 4 | **8** | **8** | 8 |
| Summarization | CodeT5-base | Teacher | 892 MB | 220M | 12 | 3072 | 768 | 12 |
| | CodeT5-small | | 242 MB | 60M | 8 | 2048 | 512 | **6** |
| | CTT-based Model | Student | **65 MB** | **17.1M** | 6 | **512** | **128** | 8 |
| Generation | CodeGen-350M | Teacher | 1.4 GB | 356.7M | 16 | 4096 | 1024 | 20 |
| | | CTT-based Model | **52.5 MB** | **~13.75M** | **4** | **128** | **128** | **6** |
| | Qwen Coder 2.5 1.5B Instruct | Teacher | 3.1 GB | 1.5B | 12 | 8960 | 1536 | 28 |
| | | CTT-based Model | **216.2 MB** | **~133.75M** | **8** | **1024** | **512** | **16** |

*4.1.1 Code Clone Detection.* In the task of CCD, we compared the student model's performance with that of the large teacher model (UniXCoder, 504 MB) and several widely adopted large-scale baselines in Table 2. Our student model, with an extremely reduced size (approximately 300× or 99.68% smaller than UniXCoder), retains 98% of its performance, highlighting that model compactness does not compromise detection quality. Additionally, the student model outperforms CodeBERT and CodeT5 in recall while matches them with CodeT5+ and Compressor.

Table 2. Precision and Recall comparison for CCD

| **Model** | **Size** | **Precision** | **Recall** |
|---|---|---|---|
| CodeBERT | 481 MB | 0.947 | 0.934 |
| CodeT5 | 892 MB | 0.95 | 0.947 |
| CodeT5+ | 446 MB | 0.94 | 0.96 |
| UniXCoder (Teacher) | 504 MB | 0.93 | 0.976 |
| Compressor | 3.1 MB | 0.935 | 0.96 |
| Avatar | 5.5 MB | 0.71 | 0.94 |
| CTT-based model | 1.6 MB | 0.93 | 0.96 |

*4.1.2 Code Summarization.* In terms of summarization, CodeT5-base-multi-sum was selected as teacher model, with CodeT5-small for a smaller variant. Results in Table 3 show that although the CTT-based student is reduced by more than 90% in size, it still maintains around 89% of the teacher's performance in terms of rouge-l, while also outperforming CodeT5-small across BLEU, METEOR, and ROUGE-L. This demonstrates that our pipeline effectively preserves summarization quality even under aggressive compression.

*4.1.3 Code Generation.* For the code generation task, we evaluate student models distilled from two teacher families: CodeGen-350M and Qwen Coder 2.5 1.5B Instruct. As shown in Table 4, the CTT-based student derived from CodeGen-350M is reduced by nearly 97% in size while maintaining similar performance to the teacher. Similarly, the Qwen-based student compresses the teacher by almost 93% yet retains a substantial fraction of performance. Table 5 further reports functional correctness of MBPP sanitized data using pass@1. Here, the CodeGen-derived student achieves 9.6% versus 15.2% for its teacher, while the Qwen-derived student records 39.5%.

Table 3. BLEU and Model Size on Code Summarization

| Model | BLEU | METEOR | ROUGE-L |
|---|---|---|---|
| CodeT5-base-multi-sum (Teacher) | 19.5 | 32.15 | 34.37 |
| CodeT5-small | 11.17 | 25.59 | 27.56 |
| CTT-based model | 15.2 | 27.21 | 30.44 |

Table 4. Performance on CodeSearchNet Challenge

| Model Family | Variant | Size | BLEU | CodeBLEU |
|---|---|---|---|---|
| Codegen-350M | Teacher | 1.4 GB | 22 | 33 |
|  | CTT-based model (Student) | 52.5 MB | 20 | 30 |
| Qwen Coder 2.5 1.5B Instruct | Teacher | 3.1 GB | 35.0 | 25.7 |
|  | CTT-based model (Student) | 216.2 MB | 28.1 | 20.64 |

Table 5. Pass@1 test on MBPP-sanitized Test set

| Model Family | Variant | Size | pass@1 test % |
|---|---|---|---|
| Codegen-350M | Teacher | 1.4 GB | 15.2 |
|  | CTT-based model (Student) | 52.5 MB | 9.6 |
| Qwen Coder 2.5 1.5B Instruct | Teacher | 3.1 GB | 58.0 |
|  | CTT-based model (Student) | 216.2 MB | 39.47 |

**Answer to RQ1**

The student model preserves almost 98% of the teacher's performance in CCD, around 89% in summarization and for code generation, textual metrics retain up to 91% of the teacher, while pass@1 reaches 68% of the teacher despite up to 97–99% reduction in size.

### 4.2 RQ2: Inference speed

After establishing comparable task performance, we next investigated the inference speed of our student model across all tasks (Figure 3).

*4.2.1 Code Clone Detection.* In CCD on GPU, CTT-based model completes inference in just 11.41 seconds. Compared to the teacher model UniXCoder, it is nearly 5× faster and completes the task 77% more quickly. Similar improvements are observed against other large models, highest for CodeT5 family (8×, 87%). On the CPU, the advantage becomes even more evident. When compared to UniXCoder, CTT-based model is 8× faster (87% reduction). Similar to GPU, improvements also noticable for other models in CPU, highest for the CodeT5+.

*4.2.2 Code Summarization.* For code summarization, CTT-based model offers significant inference acceleration. On GPU, it completes summarization in 45 seconds, delivering a 3× speedup and a 63% improvement over the teacher while 10% quicker than CodeT5-small. On CPU, CTT-based model provides 56% improvement over CodeT5-base.

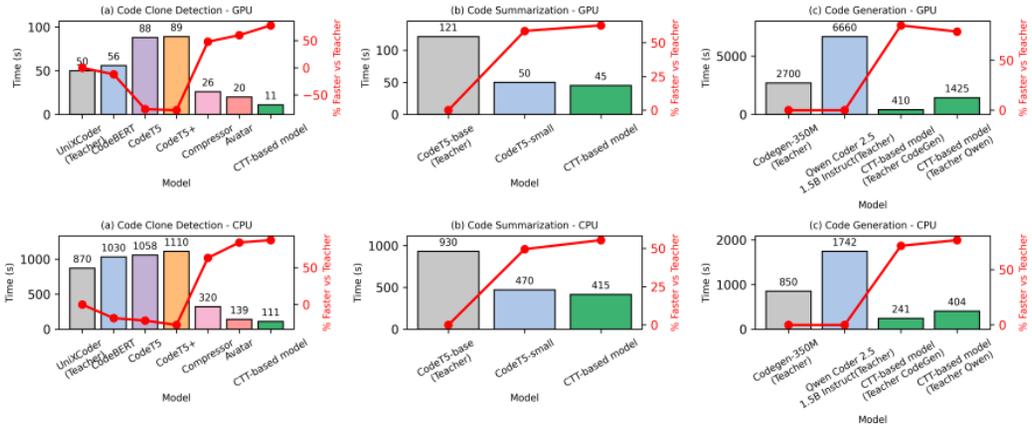

Fig. 3. Inference time and relative speed-up comparison. Red line indicates improvement % relative to teacher model.

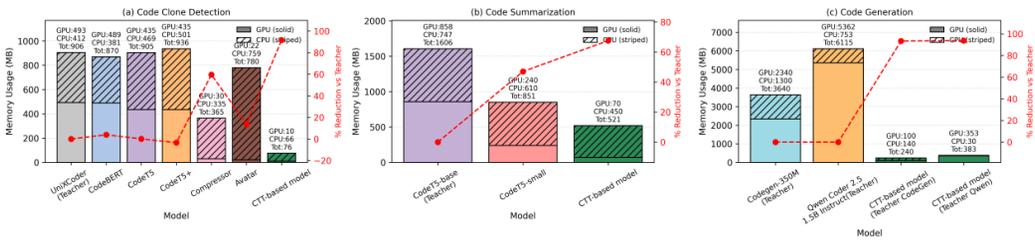

Fig. 4. Memory usage and relative reduction comparison. Each stacked bar shows GPU (solid) and CPU (striped) memory usage. Red line indicates % reduction in total memory usage relative to teacher model.

*4.2.3 Code Generation.* For code generation, we compared students and teachers within each model family. The CodeGen-derived student shows more than a 6-fold speedup on GPU and a 3-fold gain on CPU. The Qwen-derived student also delivers substantial improvements, with nearly 5-fold faster inference on GPU and 4-fold on CPU.

> **Answer to RQ2**
>
> CTT-based students achieve up to 8× GPU and 10× CPU speedups in CCD, more than 2× faster inference in summarization, and between 3–6× acceleration in code generation.

## 4.3 RQ3: Memory efficiency

Beyond task performance and inference speed, memory efficiency plays a critical role in enabling practical deployment of large models on resource-constrained devices. We analyze how much memory overhead can be reduced by compressing the teacher models into our compact student model, across all three tasks and both GPU and CPU (Figure 4).

*4.3.1 Code Clone Detection.* In the CCD, CTT-based model demonstrates substantial memory savings. On GPU, it uses 97.96% less memory than UniXCoder and shows similar reductions against others. On CPU, it reduces memory consumption by 84% compared to UniXCoder and up to 86.8% compared to CodeT5+. While Avatar is highly GPU-efficient and uses less overall memory than larger LLMs, its higher CPU memory usage is mainly due to a large intermediate size, CPU-side retention, and classifier overhead.

*4.3.2 Code Summarization.* For this task, CTT-based model yields 92% and 71% memory reduction than CodeT5-base and CodeT5-small respectively. On CPU, CTT-based model showed 40% reduction compared to CodeT5-base-multi-sum.

*4.3.3 Code Generation.* In generation, CodeGen-derived student reduces GPU memory by more than 95% and CPU usage by nearly 90%. The Qwen-derived student achieves similar savings, cutting GPU memory by over 90% and CPU memory by more than 95%.

> **Answer to RQ3**
>
> CTT-based models reduce memory usage by up to 49× on GPU and 8× on CPU in CCD, achieve more than 90% GPU savings and nearly half CPU savings in summarization, and cut GPU and CPU usage by over 90% in code generation.

## 4.4 Environmental Impact and $CO_2$ Reduction

Beyond performance and efficiency, the CTT-based model delivers a significant environmental benefit by reducing energy consumption and $CO_2$ emissions during inference. While training carbon costs are well documented, large-scale inference, especially in CI systems, developer tools, and production environments, can surpass training in total energy impact. Our results show that inference-time efficiency is not just an operational gain but an environmental imperative. To quantify emissions, we adopt the widely used CodeCarbon library [9], which accounts for CPU, GPU, and memory usage to provide a system-level estimate. Experiments were conducted in Saskatchewan (CAN), where the regional carbon intensity is 170.043 $gCO_2$/kWh. All teacher and student models were evaluated on the same machine under identical settings, so the relative $CO_2$ reductions reported are unaffected by the underlying regional carbon factor. Power consumption was sampled following CodeCarbon's default 15-second interval, with the tracker wrapping the full inference run. Table 6 reports total inference-time $CO_2$ emissions across the full evaluation for all tasks.

Across all tasks, the CTT-based students emit substantially less $CO_2$ than their teacher and baseline counterparts. In code clone detection, established models such as UniXCoder and CodeT5+ emit over twice as much as the CTT-based student, with CodeT5+ producing more than 150% additional emissions. A similar pattern emerges in summarization, where CodeT5-base emits around 68% more than the student, while even the lighter CodeT5-small incurs higher emissions despite its smaller size. The gap is most pronounced in code generation: CodeGen-350M releases over four times the emissions of its compressed student, and Qwen Coder 2.5 1.5B Instruct still produces more than double. These results reflect consistent gains across all tasks and hardware configurations of CTT. Three key factors drive this impact: (1) shorter inference durations, (2) reduced compute utilization, and (3) hardware-aware design. Together, they yield lower energy demands. In high-frequency SE workflows such as CI/CD pipelines or AI-assisted IDEs, baseline models would incur vastly higher emissions. In contrast, our CTT-based model makes scalable, low-carbon inference feasible. These outcomes align directly with the goals of the Green AI movement, addressing industry concerns from OpenAI, Google, and Microsoft about AI's environmental footprint [46]. Our approach offers an efficient model that runs on both GPU and CPU. Crucially, these environmental benefits are achieved without huge sacrificing accuracy or scalability. In sum, the CTT-based approach enables sustainable, high-frequency inference in real-world SE applications; suggesting that efficiency-driven model design is not only technically effective but essential for building responsible AI systems.

Table 6. Inference-time $CO_2$ emissions measured with CodeCarbon. The rightmost column reports how much *more* $CO_2$ a model emits compared to the CTT-based student for that task.

| Task | Model | Variant | Size | $CO_2$ (kg) | Extra vs CTT |
|---|---|---|---|---|---|
| Code Clone Detection | CTT-based model | Student | **1.6 MB** | **0.001975** | – |
| | UniXCoder | Teacher | 504 MB | 0.004213 | ▲ 113.3% |
| | CodeBERT | – | 481 MB | 0.004226 | ▲ 114.0% |
| | CodeT5 | – | 892 MB | 0.004990 | ▲ 152.7% |
| | CodeT5+ | – | 446 MB | 0.005024 | ▲ 154.4% |
| | Avatar | – | 5.5 MB | 0.002541 | ▲ 28.7% |
| | Compressor | – | 3.1 MB | 0.003176 | ▲ 60.8% |
| Code Summarization | CTT-based model | Student | **65 MB** | **0.005490** | – |
| | CodeT5-base-multi-sum | Teacher | 892 MB | 0.009356 | ▲ 70.4% |
| | CodeT5-small | – | 242 MB | 0.005953 | ▲ 8.4% |
| Code Generation | CodeGen-350M | Teacher | 1.4 GB | 0.034546 | ▲ 416.4% |
| | CTT-based model (CodeGen-350M) | Student | **52.5 MB** | **0.006690** | – |
| | Qwen Coder 2.5 1.5B Instruct | Teacher | 3.1 GB | 0.042995 | ▲ 226.3% |
| | CTT-based model (Qwen 2.5) | Student | **216.2 MB** | **0.013176** | – |

▲ indicates the percentage by which the listed model emits *more* $CO_2$ than CTT-based model.

## 5 Ablation Studies

The previous section has demonstrated the significant efficiency and performance gains of our final CTT pipeline. However, two critical questions remain: is the specific architecture of the pipeline essential to its success, and what is the impact of every component? To answer these questions and to validate that our approach is a principled, optimized methodology rather than an ad-hoc combination of techniques, we conducted two targeted ablation studies. The first study investigates whether the order of operations (the sequencing of NAS, pruning, and quantization) affects efficiency and sustainability outcomes. The second study evaluates the contribution of each individual component in the pipeline, clarifying their distinct roles in balancing performance recovery and resource efficiency.

### 5.1 Order of Operations in the Pipeline

The CTT pipeline is composed of three core compression stages: NAS, structured pruning, and quantization, followed by a performance recovery stage using KD. The ordering of the initial compression stages dictates the final structure of the student model. An incorrect sequence could lead to suboptimal architectures or amplify quantization errors. To identify the optimal sequence, we conducted an empirical study comparing all permutations of these three foundational steps across three tasks. KD was intentionally excluded from this ordering experiment, as its role is to restore performance to a finalized model architecture. For this study, we used smaller datasets (600 entries for BigCloneBench, 50 samples of MBPP sanitized test data and 300 samples of CodeSearchNet test data) and applied moderate compression settings for each task. Our goal was to isolate the impact of ordering on the model's efficiency profile. We report P50/P95 (ms) as the median and 95th-percentile inference latencies, Time (s) as the runtime, Peak Alloc (MB) as the maximum GPU memory allocated, Peak Resv (MB) as the maximum GPU memory reserved, and $gCO_2$/smp as the average carbon emissions per sample. The results, summarized in Table 7, are interesting. Across all three SE tasks, our proposed **NAS → Pruning → Quantization** pipeline consistently delivered the most favorable balance of efficiency. While alternative orderings occasionally offered a marginal benefit in a single metric, our chosen sequence demonstrated superior or highly competitive performance across latency, memory usage, and $CO_2$ emissions. Early quantization or delayed NAS leads to higher allocator churn (+10–14% PeakAlloc in code generation) and higher latency (+4–6% total time). $CO_2$ per sample tracks latency and memory increases (up to +25%). This finding provides empirical evidence that our pipeline's structure is not arbitrary instead is the effective ordering.

Table 7. Ablation study on the ordering of compression stages across three SE tasks.

| Task | Pipeline Order | P50/P95 (ms) | Time (s) | Peak Alloc (MB) | Peak Resv (MB) | $gCO_2$/smp |
|---|---|---|---|---|---|---|
| Code Clone Detection | NAS → Pruning → Quantization | 16.0 / 28.1 | 16.0 | 816.44 | 1080.00 | 0.002 |
|  | Pruning → Quantization → NAS | 16.2 / 28.8 | 16.9 | 822.82 | 1116.00 | 0.003 |
|  | Quantization → NAS → Pruning | 15.8 / 27.8 | 15.6 | 900.45 | 1138.00 | 0.002 |
| Code Summarization | NAS → Pruning → Quantization | 286.6 / 728.7 | 110 | 1260.99 | 2414.00 | 0.020 |
|  | Pruning → Quantization → NAS | 312.5 / 746.9 | 115 | 1265.84 | 2434.00 | 0.021 |
|  | Quantization → NAS → Pruning | 307.0 / 729.1 | 110 | 1507.10 | 2434.00 | 0.021 |
| Code Generation | NAS → Pruning → Quantization | 2603.7 / 2620.6 | 130 | 2766.61 | 3006.00 | 0.120 |
|  | Pruning → Quantization → NAS | 2708.7 / 2712.5 | 136 | 2867.04 | 3122.00 | 0.125 |
|  | Quantization → NAS → Pruning | 2708.7 / 2711.8 | 136 | 2867.04 | 3094.00 | 0.125 |

### 5.2 Impact of Pipeline Components

Having established the optimal ordering, our second study investigated a more fundamental question: what is the impact of every component? To confirm that, we evaluated the impact of each component on the CCD task using BigCloneBench and UniXCoder. Applying compression techniques without the final KD stage led to a catastrophic loss in performance. While combining NAS, pruning, and quantization successfully created a model that was 5× faster (Quantization alone made inference almost twice as fast), its accuracy collapsed to approximately 52%, no better than random guessing. This demonstrates that aggressive compression, while effective for improving efficiency, destroys the fine-grained knowledge required for the task. In contrast, the full teacher model performed well (93% precision, 97% recall) but was slow and resource-intensive. The crucial finding came when applying KD as the final step. This single stage restored the compressed model's performance to 93% precision and 96% recall, nearly matching the teacher while retaining all the efficiency benefits. This proves that the CTT pipeline is not merely a collection of independent techniques but a deeply integrated system. The compression stages are responsible for the dramatic gains in speed and memory, but it is the KD stage that is indispensable for breathing life back into the model, restoring its task-specific knowledge and making it practically useful.

## 6 Discussion

*Trade-Offs Between Performance and Efficiency.* Our results demonstrate that CTT-based models achieve substantial efficiency gains while preserving good performance. Across the three evaluation tasks ($N = 3$) we measured the performance gap as Teacher−Student, giving an average of +2.1 points with a 95% bootstrap confidence interval of [+0.0, +4.3]. Statistical tests (paired $t$ [53], $t = 1.70, p = 0.23$; Wilcoxon [42], $p = 0.25$) found no significant degradation. The small gap and tight interval demonstrate that compressed models stay close to teacher performance. In practice, the significant reductions in latency and memory often outweigh minor performance shifts, making CTT ideal for deployment in resource-constrained environments. Furthermore, the CTT pipeline is not rigid; its compression level can be tuned. For applications demanding maximum performance, a more conservative compression can be applied, while scenarios tolerating minor performance shifts can benefit from more aggressive optimization. This adaptability makes CTT a versatile framework for producing efficient models tailored to specific real-world requirements.

*Impact of Compression Techniques.* CTT's gains comes from a structured pipeline where NAS, pruning, quantization, and KD play complementary roles. Rather than building a universal student, we adopt task-specific pipelines to preserve in-domain fidelity across CCD, summarization, and code generation. The pipeline generalizes across encoder-only, decoder-only, and encoder–decoder models while maintaining accuracy. Our evaluation follows standard practice by using disjoint splits from the same dataset families. This confirms that CTT retains essential capacity while delivering efficiency across diverse SE tasks.

*Training Cost vs. Inference Gains.* CTT introduces a strategic, one-time training overhead (approx. 1.5× due to NAS compare to regular finetuning or teacher-student training) to achieve significant long-term inference gains. This upfront cost is quickly offset in real-world deployments, where inference accounts for up to 90% of long-term energy usage [33, 37, 38]. By front-loading the optimization, CTT produces models with substantial downstream benefits, including up to 10× faster inference, 49× lower memory usage, and an 81% reduction in $CO_2$ emissions. This makes the trade-off highly effective for developing sustainable models for production environments.

*Justifying the Pipeline: The Cost of Architectural Overhead.* A valid question is whether the complexity of the CTT pipeline is necessary. One might argue that simpler, zero-cost alternatives such as applying quantization techniques like LLM.int8() [10] and AWQ [31] directly to a model, could offer a more pragmatic path to efficiency. To investigate this, we measured the deployment efficiency of these quantization approaches against our comprehensive CTT pipeline. We explicitly do not evaluate task performance, as a fair comparison would require finetuning, which is outside the scope of this specific analysis. The goal is to isolate the impact of architectural overhead. The results: a LLM.int8() quantized CodeT5 model is 73× slower, consumes 87× more memory and emits 46× more $CO_2$ than our CTT student on code summarization. Similarly, the AWQ Qwen Coder 2.5 1.5B instruct model is 5× larger, consumes 4× more memory and emits 3× more $CO_2$ on code generation. This demonstrates architectural size is one of the dominant factor in deployment cost. Applying quantization alone, without reducing the model's structure, fails to address the fundamental bottleneck. This experiment, therefore, serves not as a direct comparison of compression techniques, but as a justification for CTT's methodology. Also while CTT requires coordinating multiple compression stages, this systematic approach is essential for achieving deployment-ready models. Our ablation studies and this analysis confirm that simpler alternatives fail to deliver comparable efficiency gains. The pipeline's modularity allows practitioners to adapt individual components while maintaining the validated ordering principles.

*Stakeholder Benefits.* The efficiency gains from CTT benefit a wide range of SE stakeholders. Individual developers gain access to fast, on-device tools without relying on high-end hardware, democratizing ML-powered code analysis. For enterprises, integrating lightweight models into CI/CD pipelines enables high-throughput analysis with minimal infrastructure and energy costs, supporting both operational and sustainability goals. Additionally, researchers and educators can more easily download, fine-tune, and extend these models with modest resources, lowering barriers to experimentation. CTT thus promotes a more inclusive, cost-effective, and sustainable AI ecosystem in SE.

## 7 Threats To Validity

***Internal Validity.*** Internal validity concerns whether the observed improvements are causally attributable to the proposed CTT-based compression method. One potential threat arises from hyperparameter configurations, which can influence model performance and efficiency. To mitigate this, we adopt consistent hyperparameter settings across all models, aligned with prior work [13, 17, 26, 50]. Where changes were necessary particularly in architectural hyperparameters and quantization-related parameters, we ensured that these modifications were applied with empirical justification. We additionally verified the correctness of pre-trained models by comparing their baseline accuracy with those reported in prior studies [17]. This design minimizes the risk that observed differences are due to tuning artifacts or implementation inconsistencies, reinforcing the internal soundness of our conclusions. In terms of ablation studies, it is important to note that these experiments were designed for methodological validation. As such, they were conducted on smaller, representative subsets of the main datasets and with moderate compression levels, enabling computationally feasible comparisons across different configurations. Their purpose is not to replicate our full-scale results, but to reveal the underlying principles that make the CTT pipeline effective.

***External Validity.*** External validity concerns how well our findings generalize beyond the specific models and tasks evaluated. We focus on three Transformer families encoder-only, decoder-only, and encoder–decoder, chosen to reflect architectural diversity in practical SE applications. The evaluation spans classification (CCD), summarization, and generation, covering a broad range of computational and semantic demands. While not exhaustive, the consistent results across these diverse models provide strong evidence of generalizability. Also, in terms of generalization of KD, each student model is compressed from a teacher specific to its respective task. CTT does not distill a generalist teacher into a generalist student; rather, each student is trained with task-specific supervision. It ensures strong in-domain performance, which is our primary deployment goal. We further ensure transparency and real-world relevance by using open-source models and public fine-tuning protocols aligned with community norms. Also, our CTT-based pipeline is architecture-agnostic and broadly applicable, supporting its use across other Transformer-based models and tasks. As a sustainability proxy, we focus on inference-time $CO_2$ emissions, which dominate the operational phase in SE specially CI/CD deployments. Though this excludes training emissions, it reflects the most impactful stage of real-world use. Finally, our evaluation relies on public benchmarks; thus, performance on proprietary industrial codebases, which are unavailable for public research, remains an open question.

***Construct Validity.*** This concerns whether our metrics accurately capture model efficiency and performance. We use established measures: precision, recall, BLEU, METEOR, ROUGE-L, pass@1 and CodeBLEU for performance; model size and memory footprint for compression; and latency and $CO_2$ emissions for environmental impact. Inference latency and memory usage may vary due to system-level noise, so we average multiple runs. Emissions are calculated using the CodeCarbon library. These safeguards ensure consistent and reliable measurement of our core constructs.

***Limitations of Scope.*** While we successfully applied CTT to models up to 1.5B parameters, scaling beyond this size (10B+) is currently constrained by GPU memory and training cost. Our experiments already demonstrate that the pipeline can be implemented effectively for billion-scale models, underscoring its feasibility for larger architectures. Also, CTT addresses real SE deployments where budget-constrained ordering determines latency, memory, and $CO_2$ and highlights the impact of pipeline ordering under explicit deployment budgets. Quantization is a pluggable module; we instantiate 8-bit quantization to match production toolchains and ensure reproducibility. Because CTT does not dependent on any particular quantizer, specialized methods (GPTQ [15]) can be swapped in without changing the algorithmic insight or budget constraints: this composability is a design strength, not a dependency.

## 8 Related Work

### 8.1 Model Compression Techniques

Several approaches have tackled model compression for SE tasks, but most are limited in scope, either in terms of the models used, tasks evaluated, or metrics reported. Token pruning strategies like DietCode [70] and Learned Token Pruning [25] reduce inference FLOPs by removing less relevant input tokens. While DietCode improves throughput on tasks like code summarization and search, its gains remain modest and task-specific, and do not apply structural changes to the model. Pruning-based methods such as Gordon et al. [16] and Michel et al. [35] showed that substantial portions of Transformer layers or attention heads can be pruned with minimal loss in accuracy. Our method instead addresses structural and architectural redundancy through task-specific empirical analysis, enabling broader and deeper efficiency gains.

Quantization offers another option for reducing inference cost. Q8BERT [69] and DoReFa-Net [71] reduce model size and latency using 8-bit and sub-8-bit representations. While effective in NLP and vision tasks, these methods were not tailored to SE-specific models. Recent SE works like ALPINE [43] improve upon this by combining token pruning with low-overhead plug-ins, achieving up to 58.1% memory savings. However, ALPINE is still constrained to encoder-only architectures. Our approach builds upon these methods by combining quantization with pruning, architecture search, and distillation in a synergistic way that maintains task performance while substantially improving latency, memory, and energy use across broader tasks and models. Ablation studies confirm that the placement of quantization in the pipeline matters—early or late quantization can increase memory allocator churn and emissions, whereas our sequence yields the best trade-offs.

KD has been widely applied in NLP (AdaBERT [7], MiniBERT [66]) and SE (Compressor [50], Avatar [49]). Compressor uses genetic search and KD to shrink encoders for binary classification, while Avatar combines KD with SMT-guided search to reduce latency and emissions. These methods remain limited to encoder-only models. Our pipeline instead unifies NAS, redundancy-informed pruning, pre-KD quantization, and KD into a coherent system spanning classification, summarization, and generation. Importantly, we empirically validate that the ordering of these stages is a crucial determinant of the final model's efficiency. Alternative orderings increase latency and $CO_2$ emissions, while our NAS→pruning→quantization→KD sequence consistently outperforms them.

### 8.2 Parameter-Efficient Fine-Tuning (PEFT)

PEFT methods aim to reduce the cost of adapting large models to new tasks without modifying their architecture. Adapter tuning [19], Compacter [34], prefix tuning [28], BitFit [4], and LoRA [20] have been explored in NLP and SE. Ayupov et al. [3] applied adapters and LoRA to PLBART and CodeT5 for code generation tasks and observed reductions in fine-tuning cost. However, PEFT techniques still keep the entire base model in memory at inference time [12], so the overall latency, peak VRAM, and energy use remain essentially unchanged. Because the full-precision backbone must still be loaded, these methods do not reduce the carbon footprint or enable deployment on hardware with tight memory budgets. Our approach differs fundamentally by compressing the model itself, resulting in true gains at inference stages.

### 8.3 Comparison Summary

In summary, most prior efforts address only isolated dimensions of model efficiency such as pruning [16, 35], quantization [69, 71], distillation [7, 50], or PEFT [3], and typically restrict evaluation to a single Transformer family or classification tasks. Also, a few works, such as Avatar [49] and ALPINE [43], incorporate energy or carbon reporting. By contrast, CTT contributes (i) a budget-aware formulation where latency, memory, and $CO_2$ are first-class objectives, and (ii) empirical evidence that the sequence NAS→pruning→quantization→KD consistently outperforms alternatives across encoder-only, encoder–decoder, and decoder-only families. Importantly, we explicitly measure inference-time latency and memory usage on both CPU and GPU in addition to $CO_2$ emission, providing a deployment-oriented view of efficiency. Furthermore, our ablation studies demonstrate that the ordering of compression stages is not incidental but a key optimization factor, and that KD is indispensable for restoring accuracy. Taken together, these contributions establish CTT as a principled, empirically validated compression pipeline that balances performance, latency, memory, and carbon cost for sustainable AI in SE.

## 9 Conclusion

As LLMs continue to drive innovation across SE tasks, their growing size and computational demands pose serious challenges for scalability, deployment feasibility, and environmental sustainability. This paper introduced CTT, a principled compression pipeline that integrates NAS, structured pruning, quantization, and KD into a unified process. Unlike prior works, CTT consistently proves effective across encoder-only, decoder-only, and encoder–decoder models on three core SE tasks: CCD, summarization, and code generation. Across these experiments, CTT achieves up to 49× smaller memory footprint, 10× faster inference on CPU and 8× on GPU, and 81% lower $CO_2$ emissions, while retaining 98% of teacher recall in clone detection, 89% in summarization, and upto 91% in code generation (textual metrics) with 68% pass@1. These results demonstrate that aggressive, budget-aware compression can deliver compact, deployable, and environmentally aligned models without sacrificing practical effectiveness. By quantifying efficiency gains in latency, memory, and emissions, CTT contributes directly to the mission of Green AI and makes efficient LLMs viable for developer machines, CI/CD pipelines, and educational settings. Furthermore, CTT's budgeting policy and NAS→pruning→quantization→KD ordering are composable: another quantizers (GPTQ) or advanced distillers can be substituted without altering the pipeline's core insight. As future work, we plan to extend CTT to token pruning, and emerging model architectures, and to quantify cross-domain generalization. In sum, CTT demonstrates that *responsible AI in SE* can be both high-performing and sustainable, setting a strong foundation for greener SE tools.

## 10 Acknowledgements

This research is supported in part by the Natural Sciences and Engineering Research Council of Canada (NSERC) Discovery Grants program, the Canada Foundation for Innovation's John R. Evans Leaders Fund (CFI-JELF), and by the industry-stream NSERC CREATE in Software Analytics Research (SOAR).

## 11 Data Availability

The code and data are available in this link: https://github.com/srlabUsask/CTT.